\documentclass[traditabstract]{aa}
\usepackage{amssymb,amsmath,epsfig,bm,color,slashed,mathrsfs}
\usepackage{amsmath,bm}
\usepackage{txfonts}
\usepackage{natbib}
\usepackage{graphicx}
\usepackage{multirow}
\bibpunct{(}{)}{;}{a}{}{,}
\usepackage[colorlinks=true,linkcolor=blue,citecolor=cyan,urlcolor=blue]{hyperref}
\usepackage{nicefrac}
\usepackage{exscale} 
\usepackage{gensymb}
\usepackage{slashed}
\usepackage{color}
\newcommand{\be}{\begin{equation}}
\newcommand{\ee}{\end{equation}}
\newcommand{\bea}{\begin{eqnarray}}
\newcommand{\eea}{\end{eqnarray}}

\newcommand{\vecnabla}{{\bm \nabla}}
\newcommand{\vecp}{{\bm p}}
\newcommand{\vecv}{{\bm v}}
\newcommand{\vecF}{{\bm F}}

\newcommand{\vecA}{{\bm A}}

\newcommand{\vecD}{{\bm D}}

\newcommand{\vecnu}{{\bm \nu}}

\newcommand{\ie}{{\it i.e.}}

\def\apj{ApJ}%
\def\apss{Ap\&SS}%
\def\mnras{MNRAS}%
\def\prc{Phys.~Rev.~C}%
\def\prd{Phys.~Rev.~D}%
\def\ssr{Space~Sci.~Rev.}%
\definecolor{red}{rgb}{0.8,0,0}
\definecolor{violet}{rgb}{0.4,0,0.4}
\definecolor{green}{rgb}{0,0.5,0.0}
\definecolor{navy}{rgb}{0.0,0.0,0.6}
\definecolor{orange}{rgb}{0.8,0.2,0.0}
\definecolor{blue}{rgb}{0.3,0.0,0.8}

\begin{document}

\title{Rapid rotational crust-core relaxation in magnetars} 

\author{Armen Sedrakian}

\institute{Institute for Theoretical Physics, J. W. Goethe University,
D-60438  Frankfurt am Main, Germany}

\abstract{
  If a magnetar interior $B$-field exceeds $10^{15}$~G, it will unpair
  the proton superconductor in the stellar core by inducing diamagnetic
  currents that destroy the Cooper pair coherence. Then, the $P$-wave
  neutron superfluid in these non-superconducting regions will couple
  to the stellar plasma by scattering of protons off the
  quasiparticles that are confined in the cores of neutron vortices by the
  strong (nuclear) force. The dynamical timescales associated with
  this interaction span from several minutes at the crust-core
  interface to a few seconds in the deep core. We show that (a) the
  rapid crust-core coupling is incompatible with oscillation models of
  magnetars that completely decouple the core superfluid from the
  crust and (b) magnetar precession is damped by the 
  coupling of normal fluids to the superfluid core and, if observed,
  needs to be forced or continuously excited by seismic activity.
}
\keywords{magnetars, rotational dynamics, superconductivity, vorticity}
\titlerunning{Rapid rotational crust-core relaxation}

\maketitle

\section{Introduction }\label{intro}

Magnetars are a class of compact stars that exhibit powerful X-ray
and soft $\gamma$-ray outburst activity, which is attributed to the
energy release stored in their internal magnetic
fields~\citep{1995MNRAS.275..255T}. Their surface $B$-fields are
measured to be a factor of thousand stronger than the fields inferred
for rotationally powered pulsars. The interior fields of magnetars are
unknown, but could be several orders of magnitude stronger than the
surface field [for recent reviews
see~\cite{2015RPPh...78k6901T,2015SSRv..191..315M}]. For interior
fields $B_{16}\sim 1$, where $B_{16}$ is the field value in units of
$10^{16}$~G, the electromagnetic interactions become of the order of
characteristic nuclear scales ($\sim$ MeV).  As a consequence, the
$S$-wave condensate of protons in the stellar core, which contains
charged Cooper pairs with opposite spins, becomes affected by the
$B$-field.  It is eventually destroyed at the second critical field
$H_{c2}$ that is known from the theory of ordinary
superconductivity~\citep{TinkhamBook}.  This \textup{{\it \textup{unpairing effect}}
}arises because the charge of Cooper pairs is
coupled to the
electromagnetic field, which winds up the trajectories of protons in
strong magnetic fields over distances smaller than the coherence length
of a Cooper pair.

In the context of magnetars, it was shown previously that the
strong density dependence of the proton pairing gaps causes the quenching of
the superconductivity to be non-uniform. As a consequence,  intermediate field magnetars with
$0.1\le B_{16}\le 5$ are partially superconducting, while  high-field
magnetars are $B_{16} \ge 5$ are fully non-superconducting
~\citep{2015PhRvC..91c5805S,2014arXiv1403.2829S}.  If the fields are
by an order of magnitude larger, $B_{16}\ge 10$, the neutron $S$-wave
condensate is unpaired by the magnetic field because of the
paramagnetic interaction of  spins of neutrons with the
$B$-field~\citep{2015arXiv151006000S}. On the other hand, the neutron
$P$-wave superfluid, which features spin-1 Cooper pairs, is unaffected
by the magnetic fields at the fundamental level. Because it is uncharged, it
cannot show Landau diamagnetism; its paramagnetic response to a
$B$-field is non-destructive because the pairing involves neutrons
with parallel spins.

The purpose of this work is to discuss the rotational coupling of
neutron superfluid in magnetar cores in the case where the fields are
large enough to unpair proton condensate. We show that  unpairing
opens a new channel of  coupling of electron-proton plasma 
to the neutron vorticity in the core. This new channel (which is
suppressed if protons are superconducting) is the scattering of
protons off neutron quasiparticles in the vortex cores by
nuclear force. This process needs to be contrasted with the scattering
of electrons off magnetized neutron vortices by purely electromagnetic
forces, which provides an upper bound on the electron mean-free-path in a
type-I superconducting case, that is, in the absence of proton
vorticity~\citep{1988ApJ...327..723A}.

The coupling strength of the superfluid to the unpaired plasma
has important implications for the macroscopic observable
manifestations of magnetars. We give two specific examples below.
\cite{2013PhRvL.111u1102G} conducted numerical simulations of
axisymmetric, torsional, magneto-elastic oscillations of magnetars
with a superfluid core to explain the observed quasiperiodic
oscillations of these objects. In doing so, it was assumed that the
superfluid is completely decoupled in the core of the star (\ie, the
neutron and proton fluids are coupled only by gravity).  The
assumptions above require a computation of the coupling time between
the neutron superfluid and charged plasma in magnetars, in
particular, in a situation where protons form normal fluids, as
assumed by \cite{2013PhRvL.111u1102G}. A second example is the
precessional motion of magnetars, more specifically, the influence of
an interior fluid on such motions.  \cite{2007Ap&SS.308..435L} discussed
the implications of the observation of precession in ordinary neutron
stars on the state of proton superconductivity in their cores
by focusing on the incompatibility of the type II superconductivity with
free precession.  An alternative is the type I superconductivity in
the cores of neutron stars with a low magnetic field~\citep{2003PhRvL..91j1101L,2005PhRvD..71h3003S,2007PhRvC..76a5801C}. A
natural extension of this discussion to magnetars requires the
knowledge of dynamical coupling of the neutron $P$-wave superfluid,
when the proton superfluidity is quenched by magnetic fields and
protons form a normal fluid.

Current observations of magnetars do not place any significant
constraints on the structure and strength of internal magnetic fields,
although it is known that purely poloidal or toroidal configurations
are unstable to certain types of instabilities. The internal fields are
likely to contain both components, but their location is unknown: they
can penetrate deep inside the core of the star or can be confined to
its crust. We assume a constant field in the core for concreteness
when discussing the unpairing effect below, while the coupling
timescales we find are local quantities that are independent of the large-scale
structure of the field.  This assumption is motivated by the fact that
the composition of plasma does change over the core (setting aside the
possibility of phase transitions in ultra-dense matter). The density
dependence of the $B$ -field has been phenomenologically parametrized
such that the field increases as some function of the density toward
the stellar center [see, for example, \cite{2014JP_S}]. While this
dependence can be easily incorporated in our discussion, it adds little
insight without a self-consistent solution of the Einstein-Maxwell
equations~\citep{2015MNRAS.447.3785C}.

This paper is structured as follows. In Sect.~\ref{sec:unpairing} we
review the unpairing effect in the cores of magnetars and the implied
structure of superfluid and superconducting shells. The relaxation
times-scales for the coupling of the magnetar core to the crust
are computed in Sect.~\ref{sec:coupling}. We discuss the
implications of our findings in Sect.~\ref{sec:implications} and
provide a summary in Sect.~\ref{sec:conclusions}.

\section{Unpairing effect}
\label{sec:unpairing}

As is well known [see, for example, \cite{TinkhamBook}], in type
II
superconductors the Ginzburg-Landau (GL) parameter, defined as
$\kappa= \lambda/\xi_p$, where $\lambda$ is the London penetration
depth of $B$-field in a superconductor, $\xi_p$ is the coherence
length, is in the range
$
1/\sqrt{2}  < \kappa <\infty.
$ 
The magnetic field is carried by electromagnetic vortices with quantum
flux $\Phi_0 = \pi/e$ (here and below $\hbar=c=1$) if the $B$-field is
in the range between the lower and upper critical fields, that
is,
$H_{c1} \le B\le H_{c2}$. If the $B$-field is larger than $H_{c2}$ , it
unpairs the Cooper pairs and destroys the superconductivity. 

The unpairing effect in a superfluid neutron and superconducting
proton {\it \textup{mixture}} was explored within the GL theory on the basis of
the following functional ~\citep{2015PhRvC..91c5805S}
\bea
\label{GL_Functional}
\mathscr{F}
[\phi,\psi] = \mathscr{F}_n[\phi,\psi]  + \mathscr{F}_{p}[\phi,\psi] 
+\frac1{4m_p}\vert \vecD\psi\vert^2 + \frac{B^2}{8\pi},
\eea
where $\psi$ and $\phi$ are the proton and neutron condensate
wave-functions, $m_p$ is the proton mass,
$\vecD = -i\vecnabla - 2e \vecA$ is the gauge-invariant derivative,
and $\mathscr{F}_n[\phi,\psi]$ and $\mathscr{F}_p[\phi,\psi]$ are the
energy-density functionals of neutron and proton condensates. 
The proximity to $H_{c2}$ guarantees that the proton condensate 
wave-function is small and its functional can be written as a
power
series 
\bea
\mathscr{F}_{p}[\phi,\psi] = \alpha\tau \vert \psi\vert^2 +\frac{b}{2}\vert
\psi\vert^4
+b'\vert \psi\vert^2 \vert \phi\vert^2,
\eea
where $\tau = (T-T_{cp})/T_{cp}$ with $T$ being the temperature and
$T_{cp}$ the critical temperature of superconducting phase transition
of protons, $\alpha$ and $b$ are the familiar coefficients of GL
expansion, while $b'$ describes the coupling between the neutron and
proton condensates.  The equations of motions of the proton condensate
associated with the GL functional \eqref{GL_Functional} are given by
the variations $\delta \mathscr{F}[\phi,\psi] /\delta \psi =0$ and
$\delta \mathscr{F}[\phi,\psi] /\delta \vecA =0$. Close to the
critical field, the GL equations can be linearized assuming further
that $\vecA$ is locally linear in coordinates, so that the $B$-field is
locally constant.  The solution of the pair of linearized GL equations
that correspond to non-vanishing $\psi$ provide the highest possible value of
the field that is still compatible with superconductivity, which is then identified
with the upper critical field~\citep{2015PhRvC..91c5805S}
\bea
\label{eq:Hc2_1}
H_{c2} &=&\frac{\Phi_0}{2\pi\xi_p^2} \left[1 + \beta(b')\right],
\eea 
%
where 
$m_p\vert\alpha\tau\vert = (2\xi_p)^{-2}$. The critical value
of the field is enhanced by $\beta\simeq 0.2$ through the density-density
coupling between neutron and proton condensates.
\begin{figure}[t]
\begin{center}
\includegraphics[width=8cm]{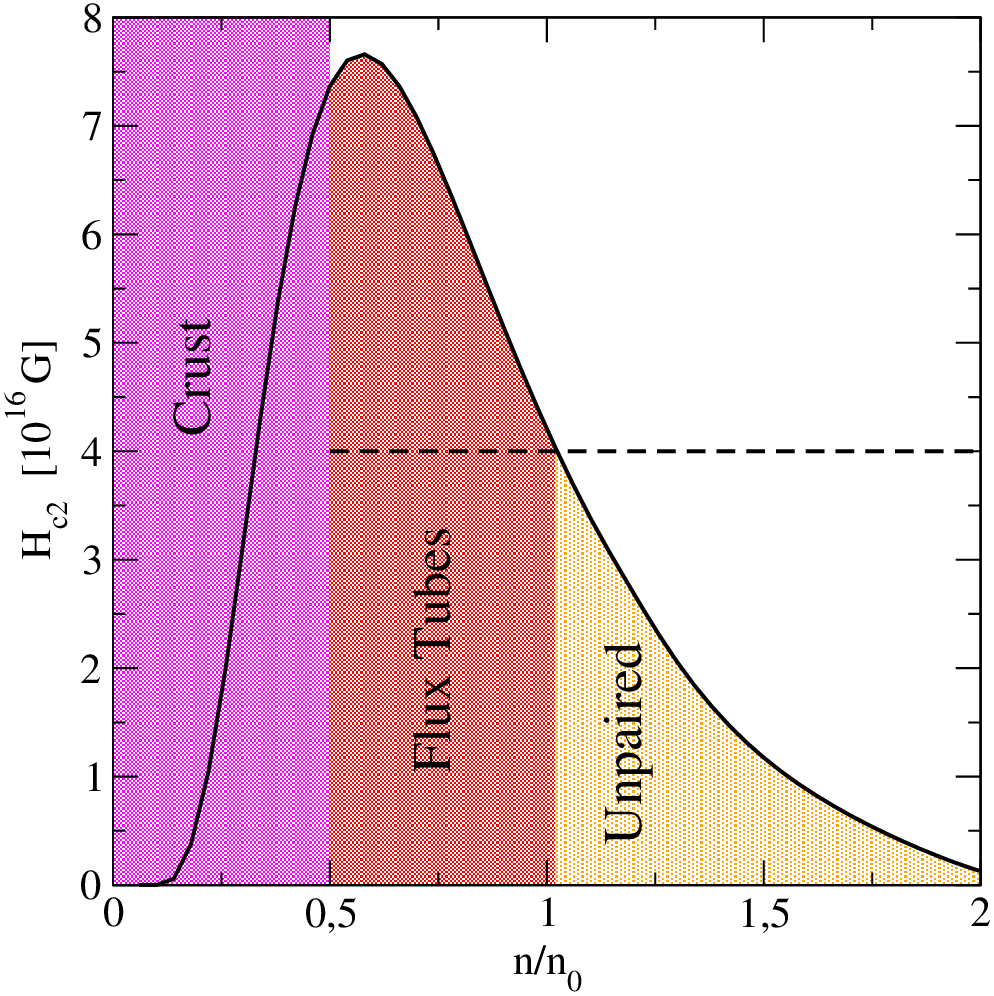}
\caption{ Dependence of the critical unpairing field $H_{c2}$ on
  baryonic density (solid line). The density is normalized to the
  nuclear saturation density $n_0= 0.16$ fm$^{-3}$ and the equation of
  state and composition of matter is the same as in
  \cite{2015PhRvC..91c5805S}. This arrangment of flux-free and
  flux-featuring phases arises in the case of intermediate field
  magnetars with average constant field $B< H_{c2}$, which is
  indicated by the dashed line. }
\label{fig:Hc2}
\end{center}
\end{figure}
The dependence of the $H_{c2}$ field on density is illustrated in
Fig.~\ref{fig:Hc2}, where we adopted the same input physics as 
described in~\cite{2015PhRvC..91c5805S}.
Independent of the details of microphysical input,
the maximum of $H_{c2} $ is attained close to the crust-core interface
(corresponding to $n_b = 0.5 n_0$, where $n_0$ is the nuclear
saturation density).  As a consequence, magnetars with
approximately constant interior fields below $H_{c2}$ will contain two
physically distinct regions: (a) the inner core, which is void of
superconductivity, and (b) the outer core, where protons are superconducting
and, consequently, proton flux-tubes (vortices) are present along
with the neutron vortex lattice induced by the rotation.  The magnetic
$B$-field in type II superconductor forms quantized electromagnetic
vortices with density $N_p = B/\Phi_0$.  These phases are enveloped by
the crust, which is threaded by non-quantized magnetic field.

\section{Rotational crust-core coupling timescales}
\label{sec:coupling}

Neutron superfluid rotates by forming an array of quantized
vortices. The areal number density of neutron vortices is given by
\bea
\label{eq:vortex_number}
N_n = \frac{2\Omega}{\omega_0}, \quad \omega_0 =  \frac{\pi}{m_n},
\eea
where $m_n$ is the bare neutron mass, $\Omega$ is the rotation
frequency of the star, and $\omega_0$ is the quantum of neutron
circulation.

Any variation in the angular velocity of the magnetar causes the free
vortices to move and leads to their new quasi-equilibrium distribution.
Hence the vortex distribution depends on their velocity field
$\vecv_L$. This velocity field is determined by the equation of motion
of a vortex, which because of the negligible vortex mass reduces to
the requirement that the sum of forces acting on its unit segment
vanishes,
\be \label{eq:force_balance}
\rho_n\omega_0 [(\vecv_S-\vecv_L)\times \vecnu] -
\eta (\vecv_L-\vecv_N) =0.
\ee
Here the first term is the Magnus force, the second is the
friction force between the vortices and the normal liquid, $\rho_n$
is the mass density of the superfluid component, $\vecv_N$ is the
velocity of the normal component, and $\eta$ is the
coordinate-dependent longitudinal (with respect to $\vecv_L- \vecv_N$)
friction coefficient.
We first consider the flux-tube free (unpaired) region and demonstrate
that it is coupled to the plasma of the star on short dynamical
coupling timescales. First we note that the non-superconducting proton
fluid will couple to the electron fluid on plasma timescales, which
are much shorter than the hydrodynamical timescales. Therefore, the
unpaired core of a magnetar can be considered as a two-fluid system
with neutron condensate forming the superfluid component and the
proton plus electron fluids forming the normal component.  Neutron
vortices (and the neutron superfluid) couples to this normal component
electromagnetically~\citep{1982PhRvD..25..967S}. However, because
protons are unpaired (\ie, excitations out of Fermi surface can be
created without the energy cost of breaking a Cooper pair), they will
scatter efficiently off neutron vortex core quasiparticles by the
nuclear force.
\begin{figure}[t]
\begin{center}
\vskip 1.cm
\includegraphics[width=9cm,height=7cm]{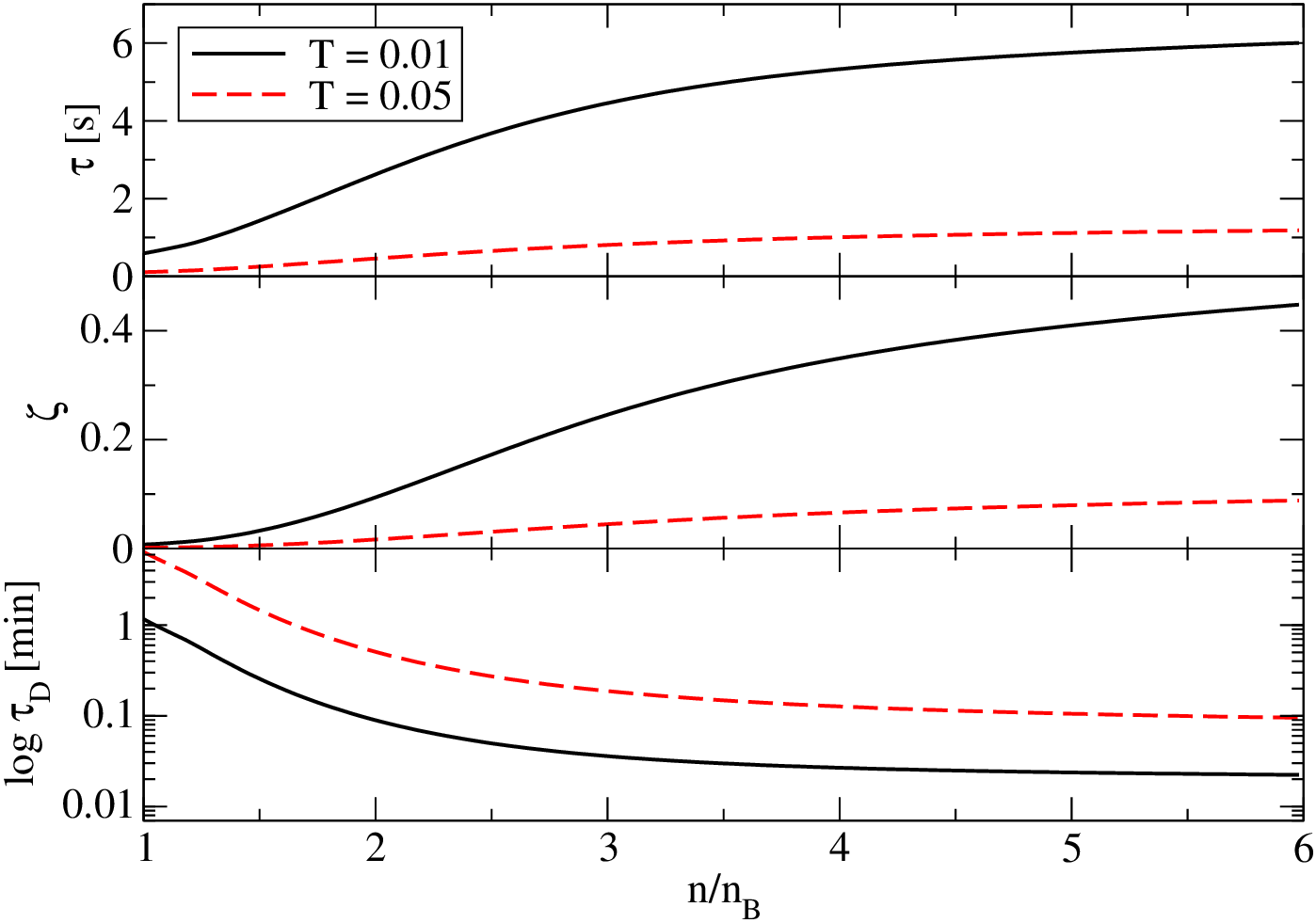}
\caption{ Relaxation timescales for protons scattering 
  off the quasiparticles that are confined in the neutron vortex core (upper 
  panel) the drag-to-lift ratio (middle panel), and dynamical 
  relaxation time (lower panel) for $T=0.01$ MeV (solid line) and 
  $T=0.05$ MeV (dashed line) and the rotation frequency $\Omega = 1$
  Hz. The results are valid in the entire density domain if the 
  constant field in the core satisfies $B>\textrm{max}\, H_{c2}$; for 
  fields lower than $H_{c2}$ , the results are valid to the right of the 
  superconducting-normal boundary (assuming the arrangement of phases
  as in Fig~\ref{fig:Hc2}). }
\label{fig:tauD}
\end{center}
\end{figure}
The solution of the Boltzmann equation for protons in the relaxation time
approximation leads to the microscopic relaxation 
timescale~\citep{1998PhRvD..58b1301S} 
\bea
\label{eq:relaxation}
\tau^{-1} = 13.81 
\frac{N_n T}{\varepsilon_{1/2}^0 m_p^*\xi_n^2}
\left(\frac{\epsilon_{Fn}}{\epsilon_{Fp}}\right)^2
e^{-\frac{\varepsilon_{1/2}^0}{T}} \frac{d\sigma}{d\Omega},
\eea
where $\epsilon_{Fn}$ and  $\epsilon_{Fp}$ are the Fermi energies of neutrons and
protons, 
$
\varepsilon_{1/2}^0 = \pi\Delta_n^2/(4\epsilon_{Fn})
$
is the lowest energy state of a neutron quasiparticle confined in the
vortex, $\Delta_n$ is the pairing gap, $d\sigma /d\Omega$ is
the differential neutron-proton scattering cross-section, and $\xi_n$ is
the neutron condensate coherence length. Here and below, for
simplicity, we do not distinguish between the neutron and proton
effective masses, meaning that we set $m_p^* = m_n^*$.

The force exerted by proton quasiparticles  {\it \textup{per
  single vortex}} is given by~\citep{1989ApJ...342..951B}
\be \label{eq:friction_force}
\vecF = \frac{2}{\tau N_n} \int f (\vecp, \vecv_L) \vecp
\frac{d^3p}{(2\pi\hbar)^3} = -\eta \vecv_L,
\ee
where $f (\vecp, \vecv_L)$ is the non-equilibrium distribution
function, which we expand assuming small perturbation about the
equilibrium distribution function $f_0$, that is ,
$
f (\vecp, \vecv_L) =
 f_0 (\vecp)+ (\partial f_0/ \partial  \epsilon)  (\vecp\cdot \vecv_L).
$
In the low-temperature limit
$\partial f_0/\partial \epsilon\simeq -\delta(\epsilon-\epsilon_{Fp})
$. The friction coefficient, after phase space integrations in 
\eqref{eq:friction_force}, is given by
\bea
 \eta &=&\frac{ m_p^* n_p}{\tau  N_n}, 
\eea 
where $n_p$ is the proton number density and $N_n$ is defined in
Eq.~\eqref{eq:vortex_number}.  The quantity characterizing the
macroscopic relaxation of superfluid is the ratio of the strengths of
viscous friction force and the Magnus force or the {\it \textup{drag-to-lift
  ratio},}
\bea
\label{eq:dragtolift}
\zeta &=& \frac{\eta}{\rho_n\omega_0}  
 = \frac{1}{2\Omega\tau}\frac{n_p}{n_n}.
\eea
Finally, the macroscopic dynamical coupling time of the superfluid to
the plasma is given by
\bea 
\label{eq:taudynamical}
\tau_D = \frac{1}{2\Omega}\left( \zeta + \zeta^{-1}\right) .
\eea
We adopt the same equation of state and nucleonic composition as in
\cite{2015PhRvC..91c5805S} to compute the numerical values of the
quantities of interest and do not repeat the details of the
input here. We do not consider hyperonic or deconfined quark degrees of
freedom. The hyperonic scattering contribution will be subdominant or
of the same order of magnitude as the proton scattering because the abundances of these species are similar in hyperon-rich matter.  The
relevant relaxation timescales in two-flavor quark matter have been
computed for color-magnetic flux tubes interacting through the Aharonov-Bohm
effect with leptons and unpaired quarks, in which case again the
strong force is involved~\citep{2010JPhG...37g5202A}.

Figure \ref{fig:tauD} shows the key results of this study: the
relaxation time \eqref{eq:relaxation}, the drag-to-lift ratio
\eqref{eq:dragtolift}, and the dynamical coupling timescale
\eqref{eq:taudynamical} as a function of baryon density in the stellar
core for $T=0.01$ and $0.05$ MeV, or equivalently for
$T= 1.2\times 10^{8}$ K and $T= 5.8\times 10^{8}$ K. An average energy
and angle-independent neutron-proton cross section $\sigma\simeq 60$
fm$^2$ and rotation frequency $\Omega = 1$~Hz were assumed. The
relaxation time increases with decreasing temperature mainly due to
the exponential Boltzmann factor in \eqref{eq:relaxation}.  The
results shown in Fig.~\ref{fig:tauD} are relevant in the entire
density range if the field in the star satisfies the condition
$B> \textrm{max} ~H_{c2}$, or in other words, if the unpairing effect acts in the
entire core. If $B < \textrm{max} ~H_{c2}$, then the results are valid
above a certain density threshold (see Fig.~\ref{fig:Hc2}). Below this
threshold density, the dynamics of the core is determined by the
vortex-flux interactions and the coupling of the electron liquid to
this conglomerate, which is not well understood.  For typical magnetar
periods of about 10 sec, that is, for spin rotations of about 1
Hz, Fig.~\ref{fig:tauD} implies that the unpaired core couples to the
plasma on short dynamical timescales, which lie in the range of
several minutes at the crust-core boundary to a few seconds deep in
the magnetar core.

\section{Implications for superfluid oscillations 
and precession}
\label{sec:implications}

We now briefly comment on some applications of the results above.
Superfluid oscillations were studied by \cite{2013PhRvL.111u1102G}
under the assumption that protons form a normal fluid (in line with
the unpairing effect, which is alluded to by these authors), but assuming
that the superfluid core is completely decoupled from the crust. This
leads to a higher Alfven speed in the core because only protons take
part in magneto-elastic oscillations and stronger penetration of these
modes inside the crust. In addition, a less massive core takes part in the
magneto-elastic oscillations, which means that the coupling to the
crust is stronger. The rapid relaxation times obtained in
Sect.~\ref{sec:coupling} imply that the two assumptions above are
incompatible because once protons are normal, they will couple the
proton-electron normal fluid to the neutron superfluid by scattering
off the neutron vortex core quasiparticles. We note that for rotation
periods characteristic of magnetars, the core is threaded by a mesh of
neutron vortices with an areal density given by
$N_n \sim 3.3\times 10^2 (\Omega/1\,\textrm{Hz})$ cm$^{-2}$ according
to Eq. \eqref{eq:vortex_number}.

In the region where $P$-wave superfluidity vanishes, the coupling
between neutron and proton fluids will be faster than our result above
by many orders of magnitude. The agreement of numerical finds of
\cite{2013PhRvL.111u1102G} with the data on oscillations of magnetars
may indicate a type II proton superconductivity and not a normal
proton fluid, in which case the coupling of neutron and proton fluids
depends on the complex ways the vortex lattices pin in the core of the
star [see, for example, ~\cite{2014ApJ...789..141L} and references
therein]. For models of magnetar oscillations in the case of type II
superconducting matter, see, for example,
\cite{2009MNRAS.396..894A,2008MNRAS.391..283V}.
\begin{figure}[t]
\begin{center}
\includegraphics[width=8cm,height=6.5cm]{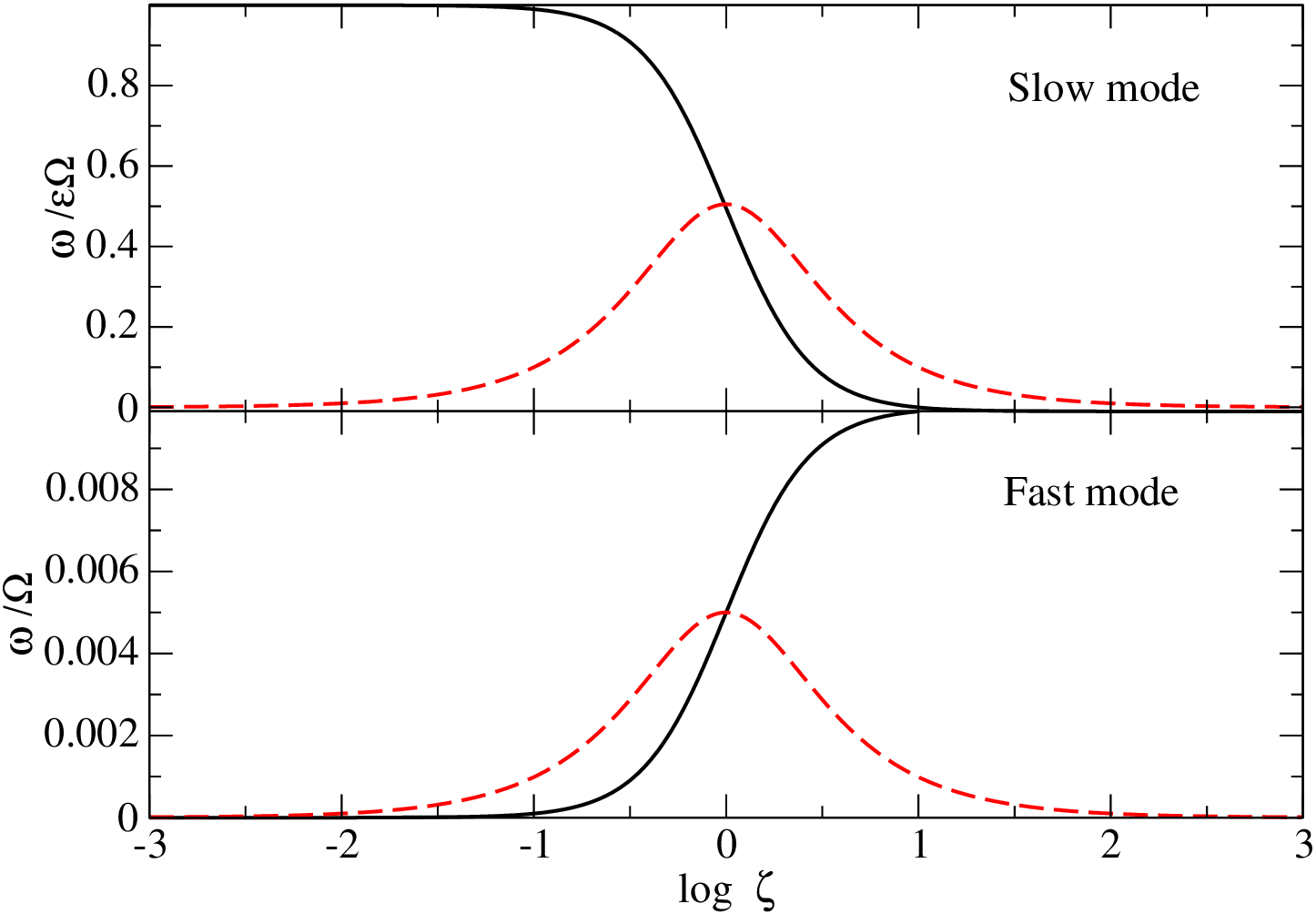}
\caption{Eigenfrequencies of precessional modes (solid lines) and
  their damping (dashed lines) of a compact star with a superfluid
  component~\citep{1999ApJ...524..341S}. The slow mode (upper panel)
  is normalized to the classical precession frequency $\epsilon\Omega$,
  where $\epsilon$ is the eccentricity.  The fast mode (lower panel)
  is a fraction of the rotational frequency. }
\label{fig:precess}
\end{center}
\end{figure}

Free precession of magnetars will be possible in a range of couplings
between the superfluid neutron fluid and the normal proton-electron
component in the core and the crust material. The eigenmodes of
precessional motion for a star with a superfluid interior were derived
for arbitrary drag-to-lift ratios in \cite{1999ApJ...524..341S} and are illustrated in Fig.~\ref{fig:precess}. Observationally
interesting are the slow precessional modes with eigenfrequency
$\sim \epsilon\Omega$, where $\epsilon$ is the eccentricity and
$\Omega$ is the rotation frequency.  This mode is precisely the
counterpart of ordinary precession in astronomical bodies. The fast
mode with eigenfrequency $\sim \Omega$ is observationally irrelevant
in the electromagnetic spectrum, but it might be an important source of
gravitational waves~\citep{2010MNRAS.402.2503J}. It is seen that for
$\zeta > 1 $ the damping of the slow mode exceeds the eigenmode
frequency, which implies that the precession is damped within a
cycle. For $\zeta \ll 1 $ precession is undamped by the superfluid
component. A comparison with the values of $\zeta$ in
Fig.~\ref{fig:tauD} shows that the low-density outer core
($\zeta\simeq 0.2$) does not affect free precession, while the
high-density inner core ($\zeta\simeq 0.4$) can cause significant
damping of precession over a cycle or so. Thus, precession of
magnetars is sensitive to the crust-coupling in the case when protons
are non-superconducting because the drag-to-lift ratios we find are
within the range of the crossover from undamped to damped
precession. Because the inner core unpairs at lower fields (see
Fig.~\ref{fig:Hc2}), we may conclude that unpairing will lead to
damping of free precession in magnetars. Our arguments apply
to free precession; it can still be observed if there is a continuous
source of excitation, such as magnetic energy, which can induce seismic
activity~\citep{2015MNRAS.449.2047L}.

\section{Summary}
\label{sec:conclusions}

The key result of this work is the demonstration that if the $B$-field
in the interior of a magnetar is large enough to unpair the proton
condensate (unpairing effect), the magnetar core will couple to the
crust on short dynamical timescales. We also computed the relevant
values of the drag-to-lift ratio that measure the influence of the
superfluid on the dynamics of normal fluid plasma. The obtained range
of this parameter lies in the region of the crossover from undamped
precession to its complete damping, therefore an observation of
precession in magnetars can shed light on the dynamical coupling
mechanism of their core to the crust. Our results indicate that
long-term precession is unlikely in magnetars and, if observed, needs
to be induced by seismic activity.  We anticipate that our results may
be useful in other contexts, such as the quasi-radial oscillations of
magnetars, their glitch and anti-glitch relaxations, and vortex shear
modes.

\section*{Acknowledgements}

The support of this research by the Deutsche
For\-schungs\-gemeinschaft (Grant No. SE 1836/3-1) and by the European
NewCompStar COST Action MP1304 is gratefully acknowledged.

\end{document}